\documentclass[prl,superscriptaddress,reprint,showpacs]{revtex4-1}

\usepackage{graphicx} 
\usepackage[usenames,dvipsnames]{xcolor}
\usepackage{tikz}
\usepackage{amsmath}
\usepackage{amssymb}
\usepackage{color}
\usepackage{float}
\usepackage{epstopdf}
\usepackage[normalem]{ulem}
\usepackage{pgfplots,pgfplotstable}
\usepackage{hyperref}
\usepackage{soul}

\usetikzlibrary[shadings]
\usetikzlibrary{shapes}
\usetikzlibrary{snakes}

\sethlcolor{white}

\pgfdeclareplotmark{ellipse}
  {\draw[draw=blue] (0,0) ellipse [x radius=1pt, y radius=2pt];}
  
  \pgfdeclareplotmark{bigellipse}
  {\draw[draw=black] (0,0) ellipse [x radius=1.5pt, y radius=3pt];}
  
    \pgfdeclareplotmark{rectangle}
  {\draw[draw=black] (0,0) rectangle (3.5pt,2.5pt);}
 
     \pgfdeclareplotmark{losange}
  {\draw[black] (-2pt,2pt) -- (-4pt,2pt);
\draw[black] (-4pt,2pt) -- (-4pt,0pt);
\draw[black] (0pt,0pt) -- (0pt,-2pt);
\draw[black] (0pt,-2pt) -- (-2pt,-2pt);
\filldraw[BlueGreen] (0,0) arc (0:90:2pt);
\filldraw[BlueGreen] (-2pt,-2pt) arc (270:180:2pt);
\filldraw[BlueGreen] (-2pt,-2pt) -- (0pt,-2pt) -- (0pt,0pt) -- (-2pt,2pt) -- (-4pt,2pt) -- (-4pt,0pt);}

\makeatletter
\providecommand*{\diff}%
        {\@ifnextchar^{\DIfF}{\DIfF^{}}}
\def\DIfF^#1{%
        \mathop{\mathrm{\mathstrut d}}%
                \nolimits^{#1}\gobblespace
}
\def\gobblespace{%
        \futurelet\diffarg\opspace}
\def\opspace{%
        \let\DiffSpace\!%
        \ifx\diffarg(%
                \let\DiffSpace\relax
        \else
                \ifx\diffarg\[%
                        \let\DiffSpace\relax
                \else
                        \ifx\diffarg\{%
                                \let\DiffSpace\relax
                        \fi\fi\fi\DiffSpace}  

{\begin{list}{}%
         {\setlength{\leftmargin}{#1}\setlength{\rightmargin}{#2}}%
         \item[]%
}
{\end{list}}

\newcommand{\vett}[1]{\mathbf{#1}}

\newcommand {\tr} {\mbox{\rm tr\,}}

{\left\lbrace\begin{array}{@{}l@{}}}%
{\end{array}\right.}

\tikzset{
precise pin/.style args={#1:#2:#3}{
pin={[%
inner sep=0pt,
pin distance=#3,
label={[%
append after command={
node[
inner sep=-1pt,
at=(\tikzlastnode.#1),
anchor=#1+180
]{#2}
}
]center:{}}
]#1:{}}
}
}

\tikzset{small dot/.style={fill=black, circle,scale=0.2}}

\begin{document}
\title{Geometry and Mechanics of Thin Growing Bilayers}
\author{Matteo Pezzulla}
\affiliation{%
Sapienza Universit\`a di Roma, via Eudossiana 18, I-00184 Roma, Italy
}%

\author{Gabriel P. Smith}
\affiliation{
Department of Mechanical Engineering, Boston University, Boston, MA, 02215.
}%

\author{Paola Nardinocchi}
\affiliation{%
Sapienza Universit\`a di Roma, via Eudossiana 18, I-00184 Roma, Italy
}%

\author{Douglas P. Holmes}
\email{dpholmes@bu.edu}
\affiliation{
Department of Mechanical Engineering, Boston University, Boston, MA, 02215.
}%

\date{\today}

\begin{abstract}
We investigate how thin sheets of arbitrary shapes morph under the isotropic in-plane expansion of their top surface, which may represent several stimuli such as nonuniform heating, local swelling and differential growth. Inspired by geometry, an analytical model is presented that rationalizes how the shape of the disk influences morphing, from the initial spherical bending to the final isometric limit. We introduce a new measure of slenderness~$\gamma$ that describes a sheet in terms of both thickness and plate shape. We find that the mean curvature of the isometric state is three fourth's the natural curvature, which we verify by numerics and experiments. We finally investigate the emergence of a preferred direction of bending in the isometric state, guided by numerical analyses. The scalability of our model suggests that it is suitable to describe the morphing of sheets spanning several orders of magnitude.
\end{abstract}


\maketitle

Cylindrically curved thin structures result from the nano-scale fabrication of semiconductor nanotubes~\cite{Tenne1992,Schmidt2001,Schmidt2001a,Stoychev2012}, and the nonuniform heating~\cite{Seffen2007a}, local swelling~\cite{Holmes2011}, and differential growth~\cite{BenAmar2005} of thin sheets. For laminated composites~\cite{Hyer1981}, electrolytic thin film deposition~\cite{Stoney1909, Freund2000}, and concrete slabs~\cite{Armaghani1987}, this cylindrical curling presents an engineering challenge, while recent work has utilized it as a mechanism for stimuli responsive self-assembly~\cite{Cho2010, Gracias2013}. The length scales of these examples range from the nanometer to the meter, suggesting that geometry dominates the deformation processes. 
Mechanically, these structures are bilayer disks in which one layer isotropically expands relative to the other. In this Letter, we show that in the asymptotic limit of large bilayer growth, any arbitrarily shaped disk will adopt a cylindrical shape whose mean curvature is three-fourth's the natural curvature.  We present an analytical model that captures both the bifurcation from spherical to cylindrical curvature~\footnote{For simple geometries, {\em e.g.} circles and ellipses, this shape evolution can be described semi-analytically by making an assumption about either the potential function or displacement field.} and the isometric limit, verified by numerics and experiments.

\begin{figure}
\begin{center}
\includegraphics[width=1\columnwidth]{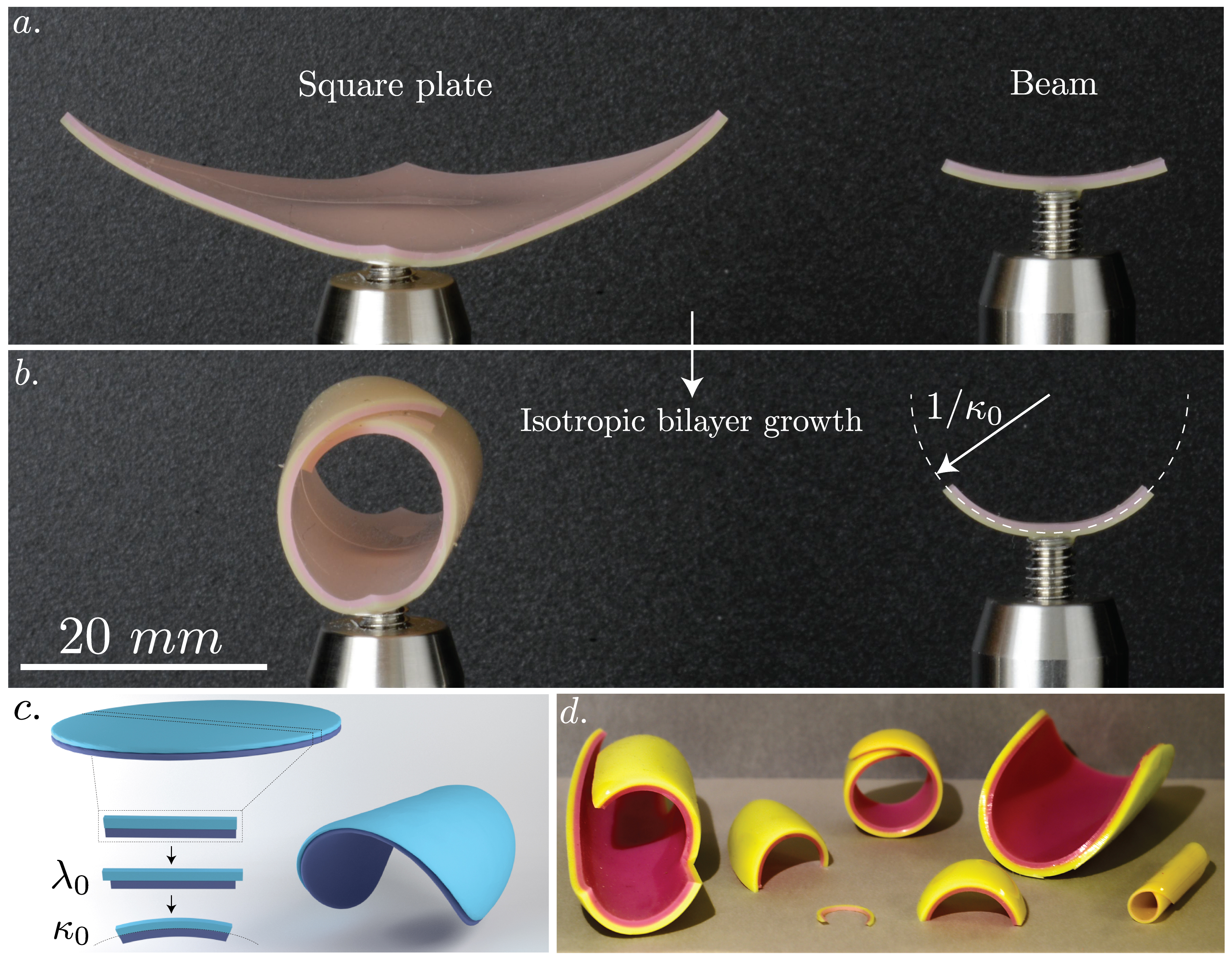}
\end{center}
\vspace{-5mm}
\caption{(a) A bilayer disk and the beam cut from it. (b) Isometric limit. (c) Measuring the natural curvature of the disk as the realized curvature of the beam. (d) Isometric states obtained by residual swelling of disks having different shapes. \label{fig-scheme}}
\end{figure}

Let us consider a thin body subjected to a dome-like (elliptic) natural curvature -- this may be thought of as the result of swelling, growth, or heating of the top surface. Within the context of non-Euclidean plates~\cite{Truesdell1952,Efrati2009}, the body may be modeled as a shell having the following first and second natural fundamental forms
\begin{equation}
\vett{\bar{a}}=
\Lambda_\textup{o}^2
\begin{pmatrix}
1&0\\
0&1\\
\end{pmatrix}\,,
\quad
\vett{\bar{b}}=
\kappa_\textup{o}
\begin{pmatrix}
1&0\\
0&1\\
\end{pmatrix}\,.
\end{equation}
The natural stretch~$\Lambda_\textup{o}$ and natural curvature~$\kappa_\textup{o}$ represent the lateral distances and curvatures that would make the sheet locally stress-free. However, as every surface in space must fulfill the Gauss-Codazzi-Mainardi equations, it is not usually possible for the sheet to realize the natural forms. Conversely, if a beam is cut from the sheet, its inherent 1D geometry will allow it to adopt a three-dimensional shape with longitudinal axis stretch and curvature equal to~$\Lambda_\textup{o}$ and~$\kappa_\textup{o}$, respectively, without any need of satisfying additional constraints. This strong connection between the sheet and the beam provides a simple means to evaluate the natural forms of the former by measuring the deformed shape of the latter [see figure~\ref{fig-scheme}~(c)].

To fully exploit this idea, we fabricated bilayer disks by spin-coating the two layers made of polyvinylsiloxane (PVS) Zhermack Elite Double 32 and 8, respectively, as shown in figure~\ref{fig-scheme}~(a-b). Free chains flow from the softer (pink) to the stiffer (green) polymer by residual swelling~\cite{Pezzulla2015}, corresponding to a dome-like natural curvature, and a conformal natural stretch. Mechanically, this corresponds to a bilayer disk where the top layer has been stretched homothetically by a factor~$\lambda^{-1}$ and then bound to the bottom layer. This results in a discontinuous three-dimensional natural metric~$\vett{\bar{g}}$ that may be approximated as~$\vett{\bar{g}}=\vett{\bar{a}}-2z\vett{\bar{b}}$, where~$z$ is the coordinate along the thickness. Approximate expressions for~$\vett{\bar{a}}$ and~$\vett{\bar{b}}$ as functions of pre-stretches were derived in~\cite{Armon2011}, albeit the natural curvature was then measured experimentally as the curvature of very small strips cut along principal directions. We overcome the lack of a predictive expression for the natural curvature by imaginarily cutting a beam from the sheet, and determining its deformed shape analytically by following~\cite{Lucantonio2014a} as~$\Lambda_\textup{o}=\Lambda_\textup{o}(\lambda,m,n)$ and~$\kappa_\textup{o}=\kappa_\textup{o}(\lambda,m,n)/h$, where~ $m$ and~$n$ are top-to-bottom thicknesses and Young moduli ratios, respectively, and~$h$ is the thickness of the sheet. Therefore, the natural forms are determined analytically as functions of~$\lambda$ and the material and geometrical ratios.

A comparison of the energies for stretching and bending suggests that a sheet exposed to such stimuli should adopt an isometric deformation in the limit of large stretch. We analytically describe the realization of this \emph{asymptotic isometry}, corresponding to a zero Gaussian curvature~$K$ and a nearly stretch-free sheet. The dimensionless energy of the shell may be written as~\cite{Armon2011}
\begin{equation}\label{eq:energy}
\begin{split}
\overline{\mathcal{U}}=&\int [(1-\nu)|\vett{a}-\vett{\bar{a}}|^2+\nu\tr^2(\vett{a}-\vett{\bar{a}})]\sqrt{|\vett{\bar{a}}|}dA+\\&+\frac{h^2}{3}\int [(1-\nu)|\vett{b}-\vett{\bar{b}}|^2+\nu\tr^2(\vett{b}-\vett{\bar{b}})]\sqrt{|\vett{\bar{a}}|}dA\,,
\end{split}
\end{equation}
where $\vett{a}$ and $\vett{b}$ are the first and second fundamental forms of the mid-surface of the deformed sheet, respectively. 
In the \emph{isometric limit}, we have~$\vett{a}=\vett{\bar{a}}$ and the stretching energy is zero. To find the deformation of the disk we then minimize the bending energy under the constraint that the mid-surface be flat, since~$\vett{a}=\vett{\bar{a}}=\Lambda_\textup{o}^2\vett{I}$.
We write~$\vett{b}$ in cartesian coordinates~$(u,v)$ so that the second fundamental form is~$Ldu^2+2Mdudv+Ndv^2$, and impose the constraint of a flat mid-surface by enforcing that its Gaussian curvature~$\Lambda_\textup{o}^{-4}(LN-M^2)$ be zero through a Lagrange multiplier.

\begin{figure}[b]
\begin{center}
\includegraphics[scale=1]{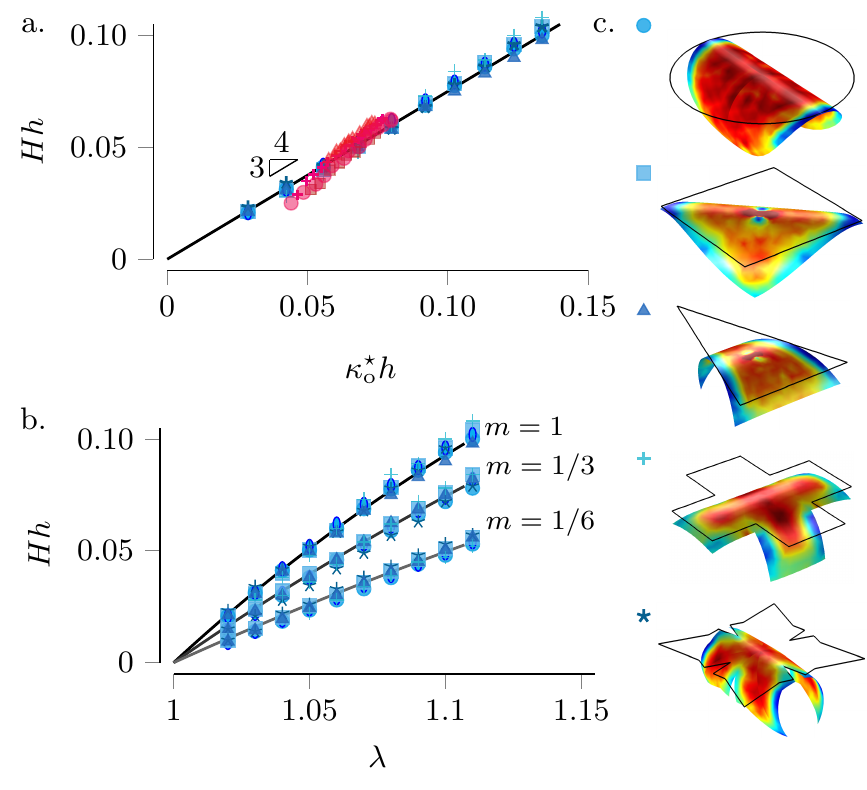}
\end{center}
\vspace{-4mm}
\caption{(a) Dimensionless mean curvature versus dimensionless rescaled natural curvature: the three-fourth law excellently captures numerical and experimental data regardless of initial shape. Numerical data (blue) refers to circle, square, leaf-like, triangle, cross and ellipse, each for different thicknesses, and experimental data (red) refers to a square, cross, triangle, and circle. (b) Dimensionless mean curvature versus $\lambda$: the isometric limit is expressed in terms of the stretch of the top layer. Three different master curves are shown, each corresponding to different thicknesses ratios~$m$. (c) Isometric shapes from COMSOL with the color code representing the mean curvature.}
\label{fig:34}
\end{figure}

For the characterization of the \emph{isometric limit}, we assume that the energy density is homogenous. Consequently, for the total energy to be minimized, it suffices to minimize its density augmented by the constraint on the Gaussian curvature imposed by the Lagrange multiplier.
The minimization gives~$L+N=\kappa_\textup{o}(1+\nu)$ and~$K=0$, a solution solely in terms of the principal invariants of the curvature tensor, and independent of the shape of the sheet: if a homogeneous energy density is assumed, no information on the bending direction can be obtained~\cite{Mansfield1965}. The result of the minimization may be rewritten in terms of the mean curvature~$H$ of the sheet as~$H=\frac{1}{2}a^{\alpha\beta}b_{\alpha\beta}=\frac{1}{2}\bar{a}^{\alpha\beta}b_{\alpha\beta}=\frac{1}{2}\Lambda_\textup{o}^{-2}(L+N)$. If the material is incompressible ($\nu=1/2$), it becomes
\begin{equation}\label{eq:34}
H=\frac{3}{4}\frac{\kappa_\textup{o}}{\Lambda_\textup{o}^2}\,,
\end{equation}
that is, in the limit of large stretch, the mean curvature of the sheet is three-fourth of the rescaled natural curvature~$\kappa_\textup{o}^\star=\kappa_\textup{o}/\Lambda_\textup{o}^2$, regardless of the shape of the sheet. Equivalently, the sheet morphs into a cylinder of radius~$R=1/(2H)$. Since~$\kappa_\textup{o}\sim1/h$ whereas~$\Lambda_\textup{o}$ does not depend on the thickness, we derive~$H\sim1/h$.

To verify our analytical prediction, we carried out experiments and numerical simulations. In the experiments, we measured the curvatures of disks of several shapes and the corresponding beams during residual swelling [figure~\ref{fig-scheme}~(a-b)] by image analysis performed using Matlab. Numerical simulations were performed to solve the geometrical problem within the context of finite (incompatible) tridimensional elasticity with large distortions using a Neo-Hookean incompressible material model~\cite{Lucantonio2014} implemented in the commercial software COMSOL Multiphysics. The sheets were made of two layers: as the top one was pre-stretched by~$\lambda^{-1}$, it was subjected to a distortion field~$\vett{F}_o=\lambda(\vett{e}_1\otimes\vett{e}_1+\vett{e}_2\otimes\vett{e}_2)+\vett{e}_3\otimes\vett{e}_3$, whereas the bottom one was not pre-stretched and therefore subjected to~$\vett{F}_o=\vett{I}$. We simulated several shapes and thicknesses, and verified that the mean curvature~$H$ was homogeneous apart from boundary layers near the edges. Then we plotted the product~$Hh$ as a function of the dimensionless natural curvature~$\kappa_\textup{o}^\star h$, or the stretch~$\lambda$ via the 1D analytical model, in the limit of large stretch. The numerical data referring to different thicknesses and the experimental data collapse to a single master curve thus verifying the analytical prediction that~$H\sim 1/h$ and the three-fourth law excellently predicts the mean curvature of the sheet regardless of its shape as shown in figure~\ref{fig:34}~(a). Figure~\ref{fig:34}~(b) shows how the dimensionless mean curvature varies with the stretch of the top layer and how it is affected by the thicknesses ratio~$m$, providing a design rule for bilayer sheets; all the data corresponding to different thicknesses and shapes collapses to master curves.

To predict and explain the transition that turns a spherical growth into a flat state~\cite{Mansfield1965}, we start from the energy in equation~\eqref{eq:energy} and compare the energies of the spherical and isometric states, without any {\em a priori} assumptions of a stress state or displacement. These assumptions usually rely on the consideration of an Airy stress function, which is known only for a very small number of sheet's shapes, and on the account of a linearized version of the Gauss's theorem~\cite{Mansfield1989,Freund2000,Seffen2007}.
Away from large stretches, the two principal curvatures are equal to each other and homogeneous throughout the sheet. However, the morphing into a spherical cap ($L=N$) becomes too costly for the sheet when the stretch, or the natural curvature, reaches a critical value. While for small values of~$\kappa_\textup{o}$, it is convenient for the sheet to stretch and increase its Gaussian curvature, above a critical threshold of~$\kappa_\textup{o}$, it is cheaper to morph isometrically. Assuming a metric with constant Gaussian curvature in Gaussian normal coordinates as in~\cite{Pezzulla2015}, the dimensionless stretching energy may be written as~$\mathcal{\overline{U}}_s=(1/9)\Lambda_\textup{o}^6K^2\int r^4dA$, where~$r$ is the undeformed radial coordinate, and we define~$\mathcal{S}=(2/9\int r^4dA/A)^{1/4}$ as the shape factor~\footnote{See supplementary material at  for complete derivation}. Also, as~$L=N$, the dimensionless bending energy is~$\mathcal{\overline{U}}_\textup{b}=h^2\Lambda_\textup{o}^{-2}A(L-\kappa_\textup{o})^2$, where $A$ is the undeformed area of the incompressible sheet. 
As~$K^2\simeq\Lambda_o^{-8}L^4$ before bifurcation, the total dimensionless energy for the spherical cap is~$\mathcal{\overline{U}}_\textup{bb}=(1/9)L^4\Lambda_\textup{o}^{-2}\int r^4dA+h^2A\Lambda_\textup{o}^{-2}(L-\kappa_o)^2$, while the total energy of the isometric shape is equal to~$\mathcal{\overline{U}}_\textup{ab}=(1/4)h^2A\Lambda_\textup{o}^{-2}\kappa_\textup{o}^2$, which becomes smaller than the energy of the spherical cap above a critical value of the natural curvature~\footnote{See supplementary material at  for details}. This analytical reasoning suggests that a bifurcation between the two principal curvatures must occur at a critical~${\kappa_\textup{o}}_\textup{b}$ to drive the sheet into an isometric shape.

\begin{figure}[t]
\begin{center}
\includegraphics[scale=1]{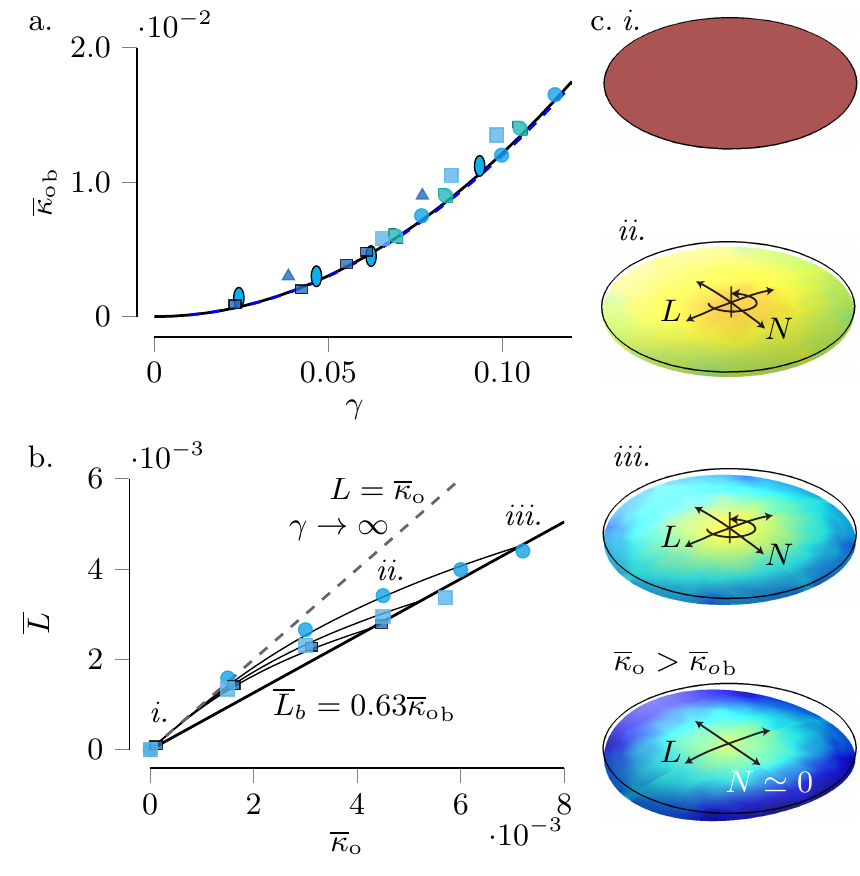}
\end{center}
\vspace{-7mm}
\caption{(a) Dimensionless natural curvature at bifurcation versus~$\gamma$: the analytical prediction given by~\eqref{eq:kbif} (solid line), in agreement with the numerical solution (dashed) of~\eqref{eq:bif}, perfectly predicts the finite element simulation results (symbols) obtained for different shapes (identified by the shape of the symbol) and thicknesses. (b) Pre-bifurcation behavior in terms of $\overline{L}(\overline{\kappa}_\textup{o})$ for different shapes (circle, square, rectangle) obtained by finite element simulations (symbols) and analytics (solid lines). The two straight lines identify the upper and lower bounds corresponding to a bending-dominated solution and the bifurcation point, respectively. (c) Deformed shapes of a thin circular disk from COMSOL.}
\label{fig:kbif}
\end{figure}
As our model does not rely on any {\em a priori} assumption on the stress state or displacements, it is suitable to study the pre-bifurcation morphing of the sheets, regardless of their shapes. Minimization of the total energy~$\mathcal{\overline{U}}_{bb}$ with respect to~$L$ yields
\begin{equation}\label{eq:prebif}
\overline{L}^3+\gamma^4(\overline{L}-\overline{\kappa}_\textup{o})=0\,,
\end{equation}
where curvatures have been nondimensionalized by~$1/h$ and denoted with bars. The geometrical dimensionless small parameter~$\gamma=h/\mathcal{S}$ encloses the shape of the sheet in the equilibrium equation and may be computed easily by its definition. The smaller~$\gamma$ is, the more slender the structure is in terms of both thickness and plate shape. Equation~\eqref{eq:prebif} can be solved analytically and it admits just one real solution~$L(\kappa_o)$ or, equivalently via the $1$D model,~$L(\lambda)$. 

When the natural curvature reaches a value such that the energy of the spherical cap coincides with the energy of the isometric state, bifurcation occurs; the equality between the two energies may be written as 
\begin{equation}\label{eq:bif}
\frac{1}{2}\overline{L}_\textup{b}^4+\gamma^4\Bigl(\overline{L}_\textup{b}^2-2\overline{L}_\textup{b}{\overline{\kappa}_\textup{o}}_\textup{b}+\frac{3}{4}{\overline{\kappa}_o}_\textup{b}^2\Bigr)=0\,,
\end{equation}
where the subscript~$b$ denotes quantities at bifurcation. This is a perturbation problem in the smallness parameter~$\gamma$ that multiplies the bending factor of equation~\eqref{eq:bif}. Seeking a perturbation expansion solution in the form~$L_b={L_b}_0+\gamma^4 {L_b}_1+O(\gamma^8)$, we obtain~$L_\textup{b}\simeq{\kappa_\textup{o}}_\textup{b}/2$, at the leading order. 
Moreover, by Taylor expanding equation~\eqref{eq:bif} in~$\gamma$ up to~$O(\gamma^{31/3})$, we derive a shape-dependent formula for the bifurcation natural curvature
\begin{equation}\label{eq:kbif}
{\overline{\kappa}_\textup{o}}_\textup{b}=\sqrt{\frac{20+14\sqrt{2}}{27}}\,\gamma^2\,.
\end{equation}
Differently from other works in the literature, this formula remarkably predicts the natural curvature at bifurcation for thin sheets having different shapes. While curvatures scale as~$1/h$ as we showed, this formula predicts that the bifurcation natural curvature scales as~$h$.
The analytical solution~$L(\kappa_\textup{o})$, evaluated at bifurcation through equation~\eqref{eq:kbif}, yields~$L_b=0.63{\kappa_o}_b$, an analytical relation that does not depend on~$\gamma$, and refines our previous asymptotic estimate. 
Figure~\ref{fig:kbif}~(a) shows the dimensionless natural curvature at bifurcation for different geometries (different~$\gamma$): the analytical model excellently predicts the threshold for different shapes and thicknesses that have been simulated in COMSOL provided that the shape is convex. The dashed blue curve corresponds to the numerical solution of equation~\eqref{eq:bif}, which justifies the asymptotic analysis that led to equation~\eqref{eq:kbif}. The model is also capable of distinguishing the pre-bifurcation behavior that sheets of different convex shapes exhibit as figure~\ref{fig:kbif}~(b) shows. For clarity, we reported the pre-bifurcation solution for three shapes. 
As~$\gamma\rightarrow\infty$, bending dominates stretching and equation~\eqref{eq:prebif} yields~$\overline{L}=\overline{\kappa}_\textup{o}$. This is also an upper-bound for all the shapes as $\overline{L}'=\gamma^4/(\gamma^4+3\overline{L}^2)$ that is maximum at~$\overline{\kappa}_\textup{o}=0$ where it is equal to~$1$ regardless of the shape: the realized curvature never exceeds the natural curvature. Also, we plotted the relation between realized and natural curvatures at bifurcation, which sets a lower-bound for~$\overline{L}$. Figure~\ref{fig:kbif}~(c) shows the numerical deformed shape of a circular disk as the stretch increases: the disk morphs from a flat state to a spherical cap before bifurcating into a cylinder. The analytical model is not capable of describing the spherical bending and the bifurcation of concave shapes, whose topological defects presumably affect their behavior~\cite{Bende2014}. Notably, the three-fourth law holds regardless of the shape, whether convex or concave.

\begin{figure}[h]
\begin{center}
\includegraphics[scale=1]{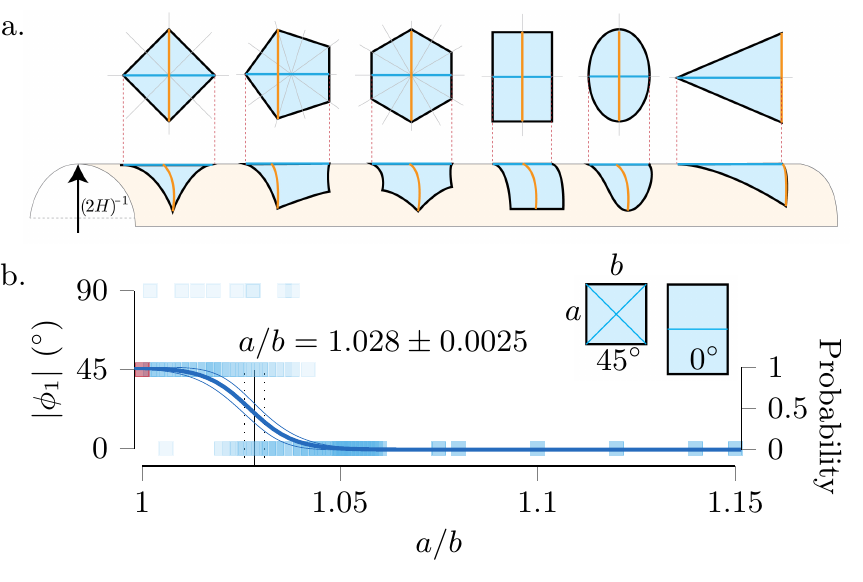}
\end{center}
\vspace{-7mm}
\caption{(a) Numerical statistics of preferred generators (blue) for different shapes. (b) Logistic analysis of the orientation data of the transition from a square to a rectangle.  \label{fig:benddirec}}
\end{figure}
Finally, we investigate the emergence of a preferred direction of bending following bifurcation by running several simulations with different meshes. While it was shown that the principal curvatures of an ellipse have no preferred orientation~\cite{Mansfield1965}, here we show that nearly all shapes select a particular direction for their generators immediately after symmetry is broken. Cylindrical generators orient themselves along an axis which preserves reflection symmetry. If multiple planes of reflection symmetry exist for a particular shape, bending will occur along the longest length of the disk (Fig.~\ref{fig:benddirec}a). For instance, a square of length $a$ and width $b$ has two unique planes of reflection symmetry, and has been observed to bend along its diagonal~\cite{Li2013}. We show with a series of simulations for $a=0.02$~m and $h=0.4$~mm that as $a/b$ is increased, and the disk transitions from square to rectangular, the orientation of the generators transitions from the diagonal to the short length $b$. Curiously, this transition does not occur immediately once $a/b>1$, as a logistic analysis of the orientation data identifies a transition at $a/b =1.028 \pm 0.0025$. A preferred bending direction in rectangular sheets has been attributed to boundary layers effects~\cite{Alben2011} that lower the energy in the disk~\cite{Fung1955}. Our results help generalize these findings for any arbitrary shape. The notion of a preferred bending direction upon bifurcation does not contradict Mansfield's classical results, as the cylindrical generators can be rotated to any arbitrary angle while remaining in equilibrium when perturbed via mechanical or chemical stimuli~\cite{Holmes2011}.

In summary, we have derived a theoretical analytical model that describes the morphing of disks subjected to dome-like natural curvatures, regardless of their shape. We also showed the emergence of an isometric limit with analytics, numerics and experiments, underlying the importance of isometries in mechanics~\cite{Dias2015}. The capability to morph 2D shapes into 3D shells by in-plane~\cite{Pezzulla2015} and transverse residual swelling opens intriguing avenues towards the precise design of soft structures.

M.P. acknowledges the National Group of Mathematical Physics (GNFM-INdAM) for support (Young Researcher Project). D.P.H. thanks D.S. Maynard for statistical analysis and acknowledges the financial support from the NSF through CMMI--1300860.

\bibliographystyle{apsrev4-1}
\bibliography{biblio}

\end{document}